\documentstyle[twoside,fleqn,espcrc2,psfig]{article}

\newcommand{\AmS}{{\protect\the\textfont2
  A\kern-.1667em\lower.5ex\hbox{M}\kern-.125emS}}

\title{ Renormalization group coupling flow of SU(3) gauge theory}

\author{ QCDTARO Collaboration\thanks{Presented by T. Takaishi} \\
Ph. de Forcrand\address{Swiss Center for Scientific Computing,
ETH-Z\"urich, CH-8093 Z\"urich, Switzerland},
T. Hashimoto\address{Department of Applied Physics, Fukui University, Fukui
910,
Japan},
S. Hioki\address{Department of Physics, Tezukayama University, Nara 631, Japan}
H. Matsufuru\address{Department of Physics, Hiroshima University, Higashi-
Hiroshima 739, Japan},
O. Miyamura$^{\rm d}$,
A. Nakamura\address{Research Institute for Information Science and Education,
Hiroshima University, Higashi-Hiroshima 739, Japan}
M.Garc\'{\i}a P\'erez\address{Institut f\"ur TheoretischePhysik,
Universit\"at Heidelberg, D69120 Heidelberg, Germany}
I.O.Stamatescu\address{FESt Heidelberg and Institut f\"ur Theoretische
Physik,
Universit\"at Heidelberg, D69120 Heidelberg, Germany},
T. Takaishi\address{Hiroshima University of Economics, Hiroshima 731-01,
Japan}
}

\begin{document}
\begin{abstract}

We present our new results on the renormalization group coupling flow obtained in 3 dimensional
coupling space $(\beta_{11},\beta_{12},\beta_{twist})$.
The value of $\beta_{twist}$ turns out to be
small and the coupling flow projected on $(\beta_{11},\beta_{12})$
plane is very similar with the previous result obtained in the 2 dimensional
coupling space.
\end{abstract}
\maketitle

\section{Introduction}
Recently a lot of studies are devoted to search for 
improved actions which have much less lattice artifacts than
the original Wilson action. Two approaches are mainly pursued to
construct the improved actions, i.e.,  Wilson's renormalization
group (RG) method\cite{Wilson}
and Symanzik's perturbation one\cite{Symanzik}.
The RG approach has been further developed by
P.Hasenfraz, Niedermayer\cite{Perfect} and
co-workers\cite{SU3}, which leads to
the classical perfect action.
Symanzik approach has become very promising in lattice simulations
with the idea of the tadpole improvement\cite{TAD}.

The classical perfect action is an approximation to the
quantum perfect action which is completely free from lattice artifacts.
Although finding quantum perfect actions is a hard task,
we have been trying to obtain them by using  the Monte Carlo
renormalization group (MCRG) method\cite{MCRG}.
In MCRG, we first block lattice configurations and then
determine renormalized coupling constants which show the
coupling flow under a certain blocking scheme.

Last year we obtained coupling flow in 2 dimensional space\cite{LAT96}.
To control truncation effects we continue our study by adding more coupling
constants.

\section{Action}
The action we use here is written as
\begin{equation}
S=\sum_i \beta_i \sum Re Tr (1-\frac{1}{3}U_i),
\end{equation}
where $i$ stands for the type of Wilson loops summarized in Table 1.

\begin{table}
\caption{Wilson loops}
\begin{tabular}{c|c} \hline
 $i$ (Type of Wilson loop) &  Path $(\nu\ne\mu\ne\rho\ne\gamma)$ \\ \hline
11  & $\nu,\mu,-\nu,-\mu$ \\
12  & $\nu,\mu,\mu,-\nu,-\mu,-\mu$ \\
22  & $\nu,\nu,\mu,\mu,-\nu,-\nu,-\mu,-\mu$ \\
Chair & $\nu,\mu,\rho,-\mu,-\nu,-\rho$ \\
Sofa  & $\nu,\mu,\rho,\rho,-\nu,-\mu,-\rho,-\rho$ \\
Twist & $\nu,\mu,\rho,-\nu,-\mu,-\rho$ \\
4Dtwist &  $\nu,\mu,\rho,\gamma,-\nu,-\mu,-\rho,-\gamma$ \\ \hline

\end{tabular}
\end{table}

Although we consider actions up to 7 coupling constants
in determining the coupling flow, here we mainly 
focus on a 3 dimensional
coupling space $(\beta_{11},\beta_{12},\beta_{twist})$,
\begin{eqnarray}
S & =& \beta_{11}\sum Re Tr (1-\frac{1}{3}U_{11})  \nonumber \\
  &  & +\beta_{12}\sum Re Tr (1-\frac{1}{3}U_{12}) \nonumber \\
  &  & +\beta_{twist}\sum Re Tr (1-\frac{1}{3}U_{twist}).
\label{eq}
\end{eqnarray}
Last year we only considered
two coupling constants $(\beta_{11},\beta_{12})$.
This year the term $\beta_{twist}$ in Eq.(\ref{eq})
is newly added.

\section{Coupling flow}
\subsection{Technique}
There are several determination techniques of coupling constants.
At the early stage of our study we used
the demon method\cite{Creutz,demon,Takaishi}
which needs an extra simulation (microcanonical simulation)
to obtain values of the coupling constants.
We now use the Schwinger-Dyson (SD) equation method\cite{SD}
which is computationally simple and  needs no extra simulation.

Truncation effects may cause different results in the two methods. 
Indeed in 2 coupling space $(\beta_{11},\beta_{12})$ 
we found $10\%$ difference in the coupling constants.
If the truncation effects would be negligibly small the result should be 
same in both cases.

\subsection{Simulation}
We employ the lattice of the size of $8^4$. 
We generate configurations at certain $\beta$ sets
$(\beta_{11},\beta_{12})$
then block the configurations.
The $\beta$ sets are chosen from the RG coupling flow obtained in
2 dimensional space in Ref.\cite{LAT96}
and the Iwasaki action. At each set of $\beta$ about 200 configurations
separated by 10 sweeps are used for the study.
On the blocked configurations we calculate values of Wilson loops
and correlation between Wilson loops, then solve the SD equations.

\subsection{Results}
First we show a result of coupling constants after one blocking starting at
$(\beta_{11},\beta_{12})=(11.0,-1.7) $. The result clearly shows an
exponential decay with the length of Wilson
loops,
which indicates that contribution of coupling constants associated with large
Wilson loops decreases rapidly. See Fig.1.

We now turn to the 3 dimensional coupling space
$(\beta_{11},\beta_{12},\beta_{twist})$.
Figs.2 and 3 show coupling flow projected on $(\beta_{11},\beta_{12})$ and
$(\beta_{11},\beta_{twist})$ coupling space, respectively.
The result projected on the  $(\beta_{11},\beta_{12})$ space is very similar
with the previous result\cite{LAT96} in 2 dimensional space
and the value of coupling constant $\beta_{twist}$ is very small compared to
other two coupling constants,
which means adding the coupling constant $\beta_{twist}$ does not
change the coupling flow very much.
Fig.4 shows the coupling flow drawn in 3 dimensional space.
The renormalized trajectory may be read off joining
the end points of the arrows.

\begin{figure}[htb]
\vspace{5pt}
\caption{$|\beta\times S|$ vs. length of Wilson loop,
where $S$ is the multiplicity of the loop when updating
one link.
}

\unitlength 1.0mm
\begin{picture}(60,60)(-10,-10)
\def\xw{60.000000} \def\yw{40.000000}
\put(13,-8){\Large length of links		}
\put(8.571428,37.327999){\circle*{1.500000}}
\put(10.571428,36.327999){\Large{$\beta_{11}$}}
\put(25.714285,29.120001){\circle*{1.500000}}
\put(17.714285,28.120001){\Large{$\beta_{12}$}}
\put(25.714285,31.120001){\circle*{1.500000}}
\put(27.714285,30.120001){\Large{$\beta_{chair}$}}
\put(25.714285,26.480000){\circle*{1.500000}}
\put(27.714285,25.480000){\Large{$\beta_{twist}$}}
\put(42.857143,7.224000){\circle*{1.500000}}
\put(44.857143,5.724000){\Large{$\beta_{22}$}}
\put(42.857143,9.696000){\circle*{1.500000}}
\put(44.857143,9.196000){\Large{$\beta_{sofa}$}}
\put(42.857143,18.480000){\circle*{1.500000}}
\put(43.857143,17.480000){\Large{$\beta_{4Dtwist}$}}
\put(0.000000,0){\line(0,1){1}}
\put(8.571428,0){\line(0,1){1}}
\put(17.142857,0){\line(0,1){1}}
\put(25.714285,0){\line(0,1){1}}
\put(34.285713,0){\line(0,1){1}}
\put(42.857143,0){\line(0,1){1}}
\put(51.428570,0){\line(0,1){1}}
\put(60.000000,0){\line(0,1){1}}
\put(-25.714285,0){\line(0,1){1}}
\put(-25.714285,0){\line(0,1){1}}
\put(0.000000,-3.5){\large 3}
\put(8.571428,-3.5){\large 4}
\put(17.142857,-3.5){\large 5}
\put(25.714285,-3.5){\large 6}
\put(34.285713,-3.5){\large 7}
\put(42.857143,-3.5){\large 8}
\put(51.428570,-3.5){\large 9}
\put(58.900000,-3.5){\large 10}
\put(0,0.000000){\line(1,0){1}}
\put(\xw,0.000000){\line(-1,0){1}}
\put(0,8.000000){\line(1,0){1}}
\put(\xw,8.000000){\line(-1,0){1}}
\put(0,16.000000){\line(1,0){1}}
\put(\xw,16.000000){\line(-1,0){1}}
\put(0,24.000000){\line(1,0){1}}
\put(\xw,24.000000){\line(-1,0){1}}
\put(0,32.000000){\line(1,0){1}}
\put(\xw,32.000000){\line(-1,0){1}}
\put(0,40.000000){\line(1,0){1}}
\put(\xw,40.000000){\line(-1,0){1}}
\put(0,0.000000){\line(1,0){1}}
\put(\xw,0.000000){\line(-1,0){1}}
\put(-8.900000,-1.500000){\large $10^{-3}$}
\put(-8.900000,6.500000){\large $10^{-2}$}
\put(-8.900000,14.500000){\large $10^{-1}$}
\put(-8.900000,22.500000){\large 1}
\put(-8.900000,30.500000){\large $10^{1}$}
\put(-8.900000,38.500000){\large $10^{2}$}
{\linethickness{0.25mm}
\put(  0,  0){\line(1,0){\xw}}
\put(  0,\yw){\line(1,0){\xw}}
\put(\xw,  0){\line(0,1){\yw}}
\put(  0,  0){\line(0,1){\yw}} }
\end{picture} 
\end{figure}
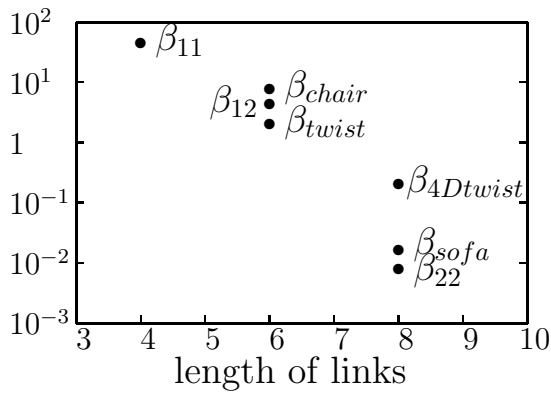

\begin{figure}[htb]
\vspace{5pt}
\centerline{\psfig{figure=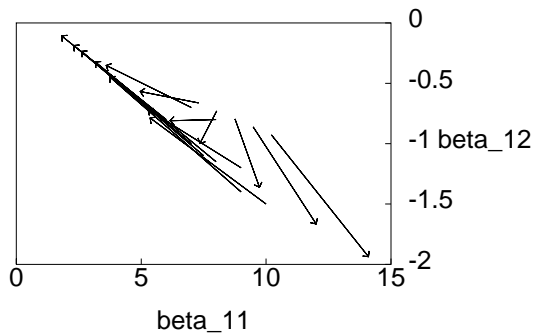,height=4cm}}
\caption{Coupling flow projected on  $(\beta_{11},\beta_{12})$ plane}
\end{figure}

\begin{figure}[htb]
\vspace{5pt}
\centerline{\psfig{figure=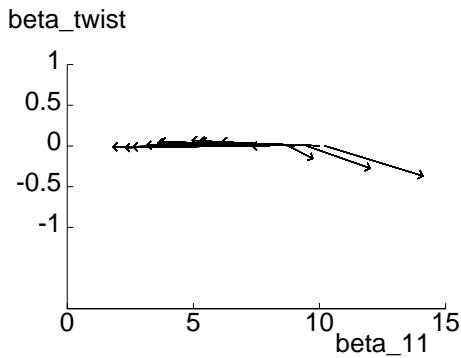,height=4cm}}
\caption{Coupling flow projected on $(\beta_{11},\beta_{twist})$  plane}
\end{figure}
\begin{figure}[htb]
\vspace{5pt}
\centerline{\psfig{figure=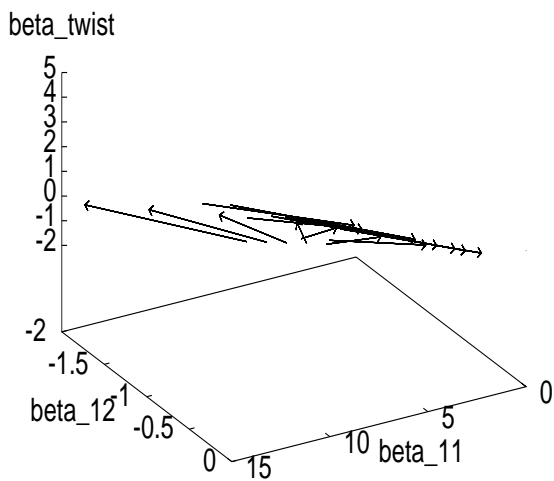,height=5.5cm,width=7cm}}
\caption{Coupling flow in 3 dimensional space}
\end{figure}

\section{Discussion}
We have studied the coupling flow in 3 dimensional coupling space.
If this coupling space is large enough to represent
the real RG flow,
the actions on the renormalized trajectory
should show the good scaling behavior.
We plan to measure several quantities on the obtained renormalized trajectory 
to check whether the actions are really "perfect".
We also prepare to extend the analysis to include $\beta_{chair}$.

\bigskip
\noindent
{\bf Acknowledgement}
The calculations reported here were done mainly on VPP500 at KEK.  We
thank the staffs at KEK for their support and hospitality.

\end{document}